\documentclass[aps,twocolumn,groupedaddress,showpacs]{revtex4}
\usepackage{longtable}
\usepackage{tabularx}
\usepackage{graphicx}
\usepackage{dcolumn}
\usepackage{bm}
\def\v_op{ \hat{\mathbf v} }

\newcommand{\ra}{\rangle}
\newcommand{\q}{{\bf {q}}}
\renewcommand{\k}{{\bf{k}}}
\newcommand{\kp}{{\bf{k'}}}

\newcommand{\nn}{\nonumber}
\newcommand{\be}{\begin{equation}}
\newcommand{\ee}{\end{equation}}
\newcommand{\bea}{\begin{eqnarray}}
\newcommand{\eea}{\end{eqnarray}}
\newcommand{\etal}{ {\it{et.~al.\,}}}

\renewcommand{\inf}{\infty}

\begin{document}

\title{A study of H+H$_2$ and several H-bonded molecules
  by phaseless auxiliary-field quantum Monte Carlo with planewave
  and Gaussian basis sets}

\author{W. A. Al-Saidi}
\altaffiliation[Present address: ]{Cornell Theory Center, Rhodes Hall,
  Cornell University, Ithaca NY 14850}
\author{Henry Krakauer}
\author{Shiwei Zhang}
\affiliation{Department of Physics, College of William and Mary, Williamsburg,
Virginia 23187-8795}

\date{\today}

\begin{abstract}

We present phaseless auxiliary-field (AF) quantum Monte Carlo (QMC)
calculations of the ground states of some hydrogen-bonded systems.
These systems were selected to test and benchmark different aspects of
the new phaseless AF QMC method.  They
include the transition state of H+H$_2$ 
near the equilibrium geometry and in the van der Walls limit, as well
as H$_2$O, OH, and H$_2$O$_2$ molecules. Most of these systems present
significant challenges for traditional independent-particle electronic
structure approaches, and many also have exact results available.  The
phaseless AF QMC method is used either with a planewave basis with
pseudopotentials or with all-electron Gaussian basis sets.  For some
systems, calculations are done with both to compare and characterize
the performance of AF QMC under different basis sets and different
Hubbard-Stratonovich decompositions.  Excellent results are
obtained using as input single Slater determinant wave functions taken
from independent-particle calculations. 
Comparisons of the Gaussian based AF QMC results with exact full configuration 
show that the errors from controlling the phase problem with the 
phaseless approximation are small.
At the large basis-size limit, the AF QMC results using both types of 
basis sets are in good
agreement 
with each other and with experimental values.

\end{abstract}

\maketitle

\section{Introduction}

Quantum Monte Carlo (QMC) methods \cite{QMC_rmp,zhang_krakauer} offer a unique way to treat
explicitly the many-electron problem. The
many-body solution is obtained in a statistical sense by
building stochastic ensembles that sample the wave function in some
representation. 
This leads to computational costs that scale as a low power with the 
number of particles and basis size.
Although in practice QMC methods
are often not exact,
they have shown considerably
greater accuracy than traditional electronic structure approaches
in a variety of systems.
They are
increasingly applied and are establishing themselves as a unique
approach for studying both realistic materials and
important model systems.

Recently, a new phaseless
auxiliary-field QMC (AF QMC) method has been developed and applied for
electronic structure calculations \cite{zhang_krakauer,gafqmc}. 
This method is formulated in a many-particle Hilbert
space whose span is defined by a single-particle basis set. The freedom to choose
the basis set can potentially result in increased  
efficiency. This can be very useful both for quantum chemistry
applications 
and in calculations with model Hamiltonians. Further, it is straightforward
in this method to exploit well-established techniques of independent-particle theories
for the chosen basis set. The ability to use any single-particle basis
is thus an attractive feature of the AF QMC method. On the other hand, the use
of finite basis sets requires monitoring the convergence of calculated 
properties and extrapolation of the results to the infinite basis size limit.

Planewave and Gaussian basis sets are the most widely used in
electronic structure calculations. Planewaves are appealing, because they
form a complete orthonormal basis set, 
and convergence with respect to basis
size is easily controlled.  
A single energy cutoff parameter $E_{{\rm cut}}$ controls the basis
size by including all planewaves with wavevector
$\k$ such that $\k^2/2 < E_{\rm{cut}}$
(Hartree atomic units are used throughout the paper).
The infinite basis limit is approached by simply increasing $E_{{\rm
cut}}$ \cite{payne_rmp}. 
Localized basis sets, by contrast, offer a compact and efficient
representation of the system's wavefunctions. Moreover, the resulting
sparsity of the Hamiltonians can be very useful in $\cal{O} (N)$
methods. Achieving basis set convergence, however,
requires more care. 
For Gaussian basis sets, quantum chemists have compiled
lists of basis sets of increasing quality for most of the elements
\cite{basis_sets_web}. Some of these basis sets have been designed
for basis extrapolation not only in mean-field theories, but also in
correlated calculations
\cite{dunning1,dunning2}.

The AF QMC method also provides 
a different route to controlling the Fermion sign problem
\cite{FermionSign,FermionSign2,Zhang,zhang_krakauer}. The 
standard diffusion Monte Carlo (DMC) method \cite{QMC_rmp,anderson_fixednode,dmc} employs the fixed-node 
approximation \cite{anderson_fixednode} 
in real coordinate space. The AF QMC method uses 
random walks in a manifold of Slater determinants (in
which antisymmetry is automatically imposed on each random walker).
The Fermion sign/phase problem is controlled approximately
according to the overlap of each random walker (Slater determinant) with a trial wave function.  
Applications of the phaseless AF QMC method to date, 
including second-row 
systems \cite{zhang_krakauer} and transition metal 
molecules \cite{alsaidi_tio_mno}
with 
planewave basis sets, and 
first-row \cite{gafqmc} and post-d \cite{gafqmc_postd}
molecular systems with Gaussian basis sets,
indicate that 
this often reduces the reliance of the results on the quality of the
trial wave function.  
For example, with single determinant trial wave functions, 
the calculated total energies 
at equilibrium
geometries in molecules
show typical systematic errors
of no more than a few milli-Hartrees compared to exact/experimental 
results. This
is roughly comparable to that of CCSD(T) (coupled-cluster with single and
double excitations plus an approximate treatment of triple
excitations).  For stretched bonds in H$_2$O \cite{gafqmc} as well as
N$_2$ and F$_2$ \cite{gafqmc_bond}, the AF QMC method exhibits better overall
accuracy and a more uniform behavior than CCSD(T) in mapping the
potential energy curve.

The key features of the AF QMC method
are thus its freedom of basis choice and control of the Fermion
sign/phase problem via a constraint in Slater determinant space.
The motivation for this study is
therefore two-fold. First, we would like to further benchmark the
planewave AF QMC method in challenging conditions, with large basis
sets and correspondingly many auxiliary fields. Here we examine the
transition state of the H$_2$+H system as well as several
hydrogen-bonded molecules.  These are relatively simple systems, which
have been difficult for standard independent-electron methods and for
which various results are available for comparison.  Secondly, we are
interested in comparing the performance of the AF QMC method using two
very different basis sets, namely, planewave basis sets together with
pseudopotentials, and all-electron Gaussian basis sets.  For this,
calculations are carried out with Gaussian basis sets for H$_2$ and
the H$_2$+H transition state, and comparisons are made with the 
planewave calculations. Additional Gaussian benchmark calculations are
carried out in collinear H$_2$+H near the Van der Waals minimum,
which requires resolution of the energy on extremely small scales. 

The rest of the paper is organized as follows. In the next section we
outline the relevant formalism of the AF QMC method. The planewave and
pseudopotential results are presented in
Sec.~\ref{sec:planewave_results}, including a study of the
dissociation energy and the potential energy curve of H$_2$, the
transition state of H$_3$, and the dissociation energies of several
hydrogen-bonded molecules.
In Sec.~\ref{sec:gaussian_results}, we use a Gaussian basis to study
the potential energy curves of H$_2$ and H+H$_2$, and compare
some of these results with the AF QMC planewave
results.  Finally, we conclude in Sec.~\ref{sec:summary} with a brief summary.

\section{AF QMC Method}
\label{sec:AFQMC-method}

The auxiliary-field quantum Monte Carlo method has been described 
elsewhere \cite{zhang_krakauer,gafqmc}. Here we outline the 
relevant formulas to facilitate the ensuing discussion. The method
shares with other QMC
methods its use of the imaginary-time propagator $e^{-\beta {\hat H}}$
to obtain the ground state $\left| \Psi_G \right\rangle$ of ${\hat
H}$:
\be 
\left| \Psi_G \right\rangle \propto \lim_{\beta \to \infty} e^{-\beta
  {\hat H}} \left| \Psi_T \right\rangle \label{eq:proj}.  
\ee
The ground state is obtained by filtering out the excited state
contributions in the trial wave function $\left| \Psi_T
\right\rangle$, 
provided that $|\Psi_T\ra$ has a non-zero overlap with 
$|\Psi_G\ra$.

The many-body electronic Hamiltonian ${\hat H}$ can be
written in any one-particle basis as,
\bea
{\hat H} &=&{\hat H_1} + {\hat H_2}; \nonumber \\
{\hat H_1}&=& \sum_{i,j,\sigma} {T_{ij} c_{i,\sigma}^\dagger  c_{j,\sigma}}; \nn \\
{\hat H_2}&=& {1\over2}
\sum_{
i,j,k,l,
\sigma,\sigma'
} {V_{ijkl} c_{i,\sigma}^\dagger c_{j,\sigma'}^\dagger c_{k,\sigma'} c_{l,\sigma}},
\label{eq:H}
\eea 
where $c_{i,\sigma}^\dagger$ and $c_{i,\sigma}$ are the corresponding
creation and annihilation operators of an electron with spin $\sigma$
in the $i$-th  orbital (size of single-particle basis is $M$).  The
one-electron and two-electron matrix elements ($T_{ij}$ and $V_{ijkl}$)
depend on the chosen basis, and are assumed to be spin-independent.

Equation.~(\ref{eq:proj})
is realized iteratively with a small time-step $\tau$ such that
$\beta=N\,\tau$, and the $\beta \to \infty$ limit is realized by letting $N \to \infty$. In this case, the Trotter decomposition
of  the propagator $e^{-\tau {\hat H}}$:
$
e^{-\tau {\hat H}}\doteq e^{-\tau {\hat H_1}/2} e^{-\tau {\hat H_2}}
e^{-\tau {\hat H_1}/2}+{\cal{O}}(\tau^3)
$
leads to Trotter time-step errors, 
which can be removed by extrapolation, using separate calculations 
with different values  $\tau$.

The central idea in the AF QMC method is the use of the
Hubbard-Stratonovich (HS) transformation \cite{HS}:
\begin{equation}
   e^{-\tau{\hat H_2}}
= \prod_\alpha \Bigg({1\over \sqrt{2\pi}}\int_{-\infty}^\infty
d\sigma_\alpha \,
            e^{-\frac{1}{2} \sigma_\alpha^2}
           e^{\sqrt{\tau}\,\sigma_\alpha\,
\sqrt{\zeta_\alpha}\,{\hat v_\alpha}} \Bigg), 
\label{eq:HStrans1}
\end{equation}
to map the many-body problem exemplified in $\hat{H}_2$ onto a linear
combination of single-particle problems using only {\emph{one-body
operators}} ${\hat v_\alpha}$. The full many-body interaction is
recovered exactly through the interaction between the one-body
operators $\{ {\hat v_\alpha}\}$, and all of the external auxiliary fields
$\{\sigma_\alpha\}$.   This map relies on writing the two-body
operator in a quadratic form, such as
\be
{\hat H_2} = - {1\over 2}\sum_\alpha \zeta_\alpha {\hat v_\alpha}^2, \label{eq:HStrans2}
\ee
with $\zeta_\alpha$ a real number. This can always be done, as we illustrate below using
first a planewave basis, and then any basis set. 

In a planewave basis set, the electron-electron interaction operator $\hat{H}_2$ can be written as: 
\bea
\label{eq:H2_rho}
\hat{H}_2 &=& \frac{1}{2 \Omega} \sum_{\k,\kp,\sigma,\sigma'}
\sum_{\q \neq 0}
 \frac{4 \,\pi\, e^2}{\q^2}\,c^{\dag}_{\k+\q,\sigma}
c^{\dag}_{\kp-\q,\sigma'} c_{\kp,\sigma'} c_{\k,\sigma} \nonumber \\ 
    &=& \frac{1}{2 \Omega} \sum_{\q > 0} \frac{4 \,\pi\, e^2}{\q^2}\, 
\left[\hat{\rho}(\q) \hat{\rho}(-\q)+ h.c.\right] + H'_{1}. 
\eea 
Here $c^{\dag}_{\k,\sigma}$ and $c_{\k,\sigma}$ are the creation and
annihilation operators of an electron with momentum $\k$ and spin
$\sigma$.
$\Omega$ is the supercell volume, $\k$ and $\kp$ are planewaves within
the cutoff radius, and the $\q$ vectors satisfy $|\k+\q|^2/2 <
E_{\rm{cut}}$. $\hat{\rho}(\q)=\sum_{\k,\sigma} c^\dag_{\k+\q,\sigma}
c_{\k,\sigma}$ is a Fourier component of the electron density operator, and $H'_{1}$
is a one-body term which arises from the reordering of the creation and annihilation operators.  
For each wavevector $\q$, 
the two-body term in the final expression in Eq.~(\ref{eq:H2_rho}) can be expressed
in terms of squares of the one-body operators proportional to
$\hat{\rho}(\q)+\hat{\rho}(-\q)$ and $\hat{\rho}(\q)- \hat{\rho}(-\q)$, which become
the one-body operators ${\hat v}_{\alpha}$ in Eq.~(\ref{eq:HStrans2}).

An explicit HS transformation can be given for any general basis as follows
(more efficient transformations may exist, however).
The two-body
interaction matrix $V_{ijkl}$ is first expressed as a Hermitian supermatrix
${\cal{V}}_{\mu[i,l],\nu[k,j]}$ where $\mu, \nu= 1,\ldots,M^2$. This
is then expressed in terms of its eigenvalues $(-\lambda_{\alpha})$
and eigenvectors $X_{\mu}^{\alpha}$: $ {\cal{V}_{\mu,\nu}}=
-\sum_{\alpha} \lambda_{\alpha} X_{\mu}^{* \alpha}\,X_{\nu}^{\alpha}
$. The two-body operator ${\hat H}_2$ of Eq.~(\ref{eq:H}) can be written as the
sum of a one-body operator ${\hat H}'_{1}$ and a two-body
operator ${\hat H}'_{2}$. The latter can be further expressed in terms of the
eigenvectors of ${\cal{V}}_{\mu,\nu}$ as 
\be {\hat H'}_2= -\frac{1}{4}\,\sum_{\alpha} \lambda_{\alpha} \left( 
{\hat \Lambda^\dagger}_{\alpha} {\hat \Lambda}_{\alpha}+{\hat
  \Lambda}_{\alpha} {\hat \Lambda^\dagger}_{\alpha}\right),
\label{eq:hs_decomp} \ee
where the one-body operators ${\hat \Lambda}_{\alpha}$ are defined as
\be
{\hat \Lambda}_{\alpha}= \sum_{i,l}X_{\mu[i,l]}^{\alpha} a^{\dag}_{i}
a_{l}. 
\ee
Similar to the planewave basis, for each non-zero eigenvalue
$\lambda_\alpha$, there are two one-body
operators ${\hat v}_{\alpha}$ 
proportional to ${\hat \Lambda}_{\alpha}+ {\hat
\Lambda}^{\dag}_{\alpha}$ and ${\hat \Lambda}_{\alpha} - {\hat
\Lambda}^{\dag}_{\alpha}$.  If the chosen basis
set is  real, then the HS transformation can be further simplified,
and the number of auxiliary fields will be equal to only the number of
non-zero eigenvalues $\lambda_\alpha$ \cite{gafqmc}.

The phaseless AF QMC method \cite{zhang_krakauer} used in this paper
controls the phase/sign problem \cite{Zhang,zhang_krakauer} in an approximate manner. The method recasts the imaginary-time path integral as
branching random walks in Slater-determinant space
\cite{Zhang}.  It uses a trial wave function
$|\Psi_T\ra$ to construct a {\em complex} importance-sampling transformation 
and to constrain the paths of the random walks.
The ground-state energy, computed with the so-called mixed estimator, is
approximate and not variational in the phaseless method.  The error
depends on $|\Psi_T\rangle$, vanishing when $|\Psi_T\rangle$ is
exact. This is the only uncontrolled error in the method, in that it cannot be eliminated
systematically.  
In applications to date, $|\Psi_T\rangle$ 
has been taken as a single Slater determinant
directly from mean-field calculations, and the systematic error is
shown to be quite small \cite{zhang_krakauer,gafqmc,gafqmc_postd,alsaidi_tio_mno}.

\begin{table}[t]
\caption{Planewave based calculations of the binding energy of H$_2$ vs. supercell
  size. 
  DFT/GGA and the phaseless AF QMC results are shown.  
  All energies are in eV, and supercell dimensions are in atomic units. 
  For comparison, the 
  all-electron GGA number is $4.568$~eV \cite{nwchem}. 
  Statistical errors
  are on the last digit and are shown in parenthesis. 
The exact theoretical value is 4.746~eV \cite{roothaan}, and the experimental
value is 4.75~eV (with zero-point energy removed). }
\begin{ruledtabular}
\begin{tabularx}{2.8 in }{p{1 in} ccc}
    supercell           & DFT/GGA     &    AF QMC         \\
\hline
11$\times$9$\times$7    & 4.283    &    4.36(1)  \\
12$\times$10$\times$9   & 4.444    &    4.57(1)   \\
14$\times$12$\times$11  & 4.511    &    4.69(1)   \\
16$\times$12$\times$11  & 4.512    &    4.70(1)  \\
22$\times$18$\times$14  & 4.530    &    4.74(2)   \\
$\inf$                  & 4.531    &               \\
\end{tabularx}
\end{ruledtabular}
\label{table_BE_H2}
\end{table}

\section{Results using planewave basis sets} \label{sec:planewave_results}

Planewaves are more suited to periodic systems and require pseudopotentials
to yield a tractable number of basis functions.
However, isolated molecules can be studied with planewaves by employing
periodic boundary conditions and large supercells,
as in standard density functional theory (DFT) calculations. 
This is  disadvantageous, because one has to
ensure that the supercells are large enough to control the spurious
interactions between the periodic images of the molecule. For a given
planewave cutoff energy $E_{{\rm cut}}$, the size of the planewave basis 
increases in proportion to the volume of the supercell. Consequently,
the computational cost for the isolated molecule tends to be higher than using
a localized basis, as we further discuss in Sec.~\ref{sec:gaussian_results}.
Although the planewave basis calculations are expensive, they are
nevertheless valuable as they show the robustness and accuracy of the
phaseless AF QMC method for extremely large basis sets (and
correspondingly many auxiliary fields).

Here we study H$_2$, H$_3$, and several other
hydrogen-bonded molecules H$_2$O, OH, and H$_2$O$_2$.  
As is well known, first-row atoms like oxygen are challenging, since they 
have strong or ``hard'' pseudopotentials and require relatively large 
planewave basis sets to achieve convergence. Even in hydrogen,
where there are no core electron states, pseudopotentials are usually used, since they 
significantly reduce the planewave basis size compared to 
treating the bare Coulomb potential of the proton.
The hydrogen and oxygen pseudopotentials are generated by the OPIUM
program \cite{rappe}, using the neutral atoms as reference
configurations. The cutoff radii used in the generation of the oxygen
pseudopotentials are 
$r_{c}(s)=1.05$ and $r_{c}(p)=1.02$ Bohr, where $s$ and $p$ correspond
to $l=0$ and $l=1$ partial waves, respectively.
For hydrogen $r_{c}(s)=0.66$ Bohr was used.
These relatively small $r_c$'s are needed for both atoms, due to the
short bondlengths in H$_2$O, and result in relatively
hard pseudopotentials.
Small $r_{c}$'s, however, generally result in pseudopotentials with better 
transferability. In
all of the studies shown below, the same 
pseudopotentials were used, even in molecules with larger bondlengths. The
$E_{\rm{cut}}$ needed with these
pseudopotentials is about $41$~Hartree. This $E_{\rm cut}$ was chosen
such that the resulting planewave basis convergence errors are less
than a few meV in DFT calculations. A roughly similar
planewave basis convergence error is expected at the AF QMC level, based
on previous applications in TiO and other systems
\cite{alsaidi_tio_mno,CPC05,cherry}. These convergence errors are much smaller than the QMC 
statistical error.

The quality of the pseudopotentials is further assessed by
comparing the pseudopotential calculations with all-electron (AE) results
using density functional methods, which tests the pseudopotentials, 
at least at the mean-field level.  In all of the cases reported in
this study, we found excellent agreement between AE and pseudopotential results, except in
cases where the non-linear core correction error
\cite{nl_core,nl_core_porezag} is important in DFT (all molecules containing
oxygen), as we discuss in Sec.~\ref{sec:planewave_hbond}.

\begin{figure}[t] 
\includegraphics[width=8.5cm]{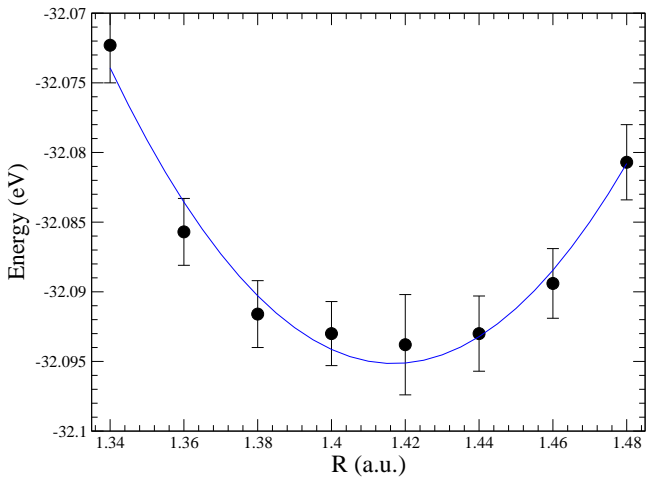} 
\caption{ The potential energy curve of H$_2$ as obtained by
  AF QMC with a planewave basis and a hydrogen pseudopotential. We
  show also a Morse potential fit for the QMC data.
  The QMC equilibrium bondlength from the fit is 1.416(4)\,Bohr to be
  compared with the exact value is 1.40083\,Bohr.  Supercell used is
  $16 \times 12 \times 11 $ Bohr$^3$. }
\label{fig_H2bond}
\end{figure}

As mentioned, AF QMC relies on a trial wave
function to control the phase problem. In the planewave calculations,
we used a single Slater determinant from a planewave based density functional
calculation obtained with a GGA functional \cite{gga}, with no
further optimizations.

\subsection{Ortho- and Para-H$_2$ molecule}

Table~\ref{table_BE_H2} summarizes results for the binding energy of the
H$_2$ molecule, using DFT/GGA and AF QMC for several supercells, and 
compares these to exact results \cite{roothaan} and experiment.
The experimental bondlength of H$_2$ was used in all the calculations.  The binding energy
is calculated as the difference in energy between the
H atom (times two) and the molecule, each placed in the same supercell. The
density functional binding energy obtained using the hydrogen
pseudopotential converges, with respect to size effect, to $4.531$~eV, 
which is in reasonable agreement with
the all-electron (i.e. using the proton's bare Coulomb potential) binding energy $4.568$~eV obtained using NWCHEM
\cite{nwchem}, and with the all-electron value of 4.540~eV reported in
Ref.~\cite{patton}. The agreement between the 
pseudopotential and all-electron results is a reflection of the good
transferability of the hydrogen pseudopotential.
The AF QMC binding energy with the largest supercell is
4.74(2)~eV, which is in excellent agreement with the experimental
value of 4.75~eV (zero point energy removed) and 
the exact calculated value of 4.746~eV 
\cite{roothaan}.

Figure~\ref{fig_H2bond} shows the H$_2$ AF QMC potential energy curve, using
a $16 \times 12 \times 11 $ Bohr$^3$ supercell.
Finite size effects, as in Table~\ref{table_BE_H2}, likely vary with 
the H$_2$ bondlength and would affect the shape of the curve.
Using a Morse potential fit, we obtained an estimated bondlength of
$1.416(4)$\,Bohr. (Using a 2nd or 4th
order polynomial fit leads to similar results; with 4th order fit the error bar is three times
larger). 
For comparison, the exact equilibrium bondlength of H$_2$ is
$1.40083$~a.u \cite{roothaan}, and the DFT/GGA bondlength is
1.4213\,Bohr. 

The energy difference between the ortho- and para-H$_2$ spin states ($^1\Sigma$ and
$^3\Sigma$, respectively) was also calculated.
We note that for the singlet H$_2$ two-electron system, a HS transformation
based on the magnetization \cite{Silvestrelli93} can be made to 
eliminate the sign problem and thus the need for the phaseless approximation.
In this case the AF QMC calculations will become exact.  
This is not done here, since our
goal is to benchmark the general algorithm.  
The calculations for both
ortho- and para-H$_2$ were at the experimental bondlength of 
ortho H$_2$. 
Table~\ref{table_H2_triplet} summarizes the results.
The exact value obtained by Kolos and Roothaan is
$10.495$~eV \cite{roothaan}, with which the AF QMC value at the larger supercell size   
is in excellent agreement.

\begin{table}[tb]
\caption{Planewave based AF QMC energies of the ortho- ($^1\Sigma$) and para-H$_2$
  ($^3\Sigma$) molecule for two supercell sizes. The bondlength was
  fixed  at   $R=1.42$ Bohr in all cases. All energies are in eV. The exact calculated energy gap $\Delta$
  is $10.495$~eV \cite{roothaan}.  
  Statistical errors are on the last digit and are shown in
  parenthesis.}
\begin{ruledtabular}
\begin{tabularx}{4 in}{p{1.1 in} cccc}
  supercell                       &  $^1\Sigma$   &$^3\Sigma$ &   $\Delta$        \\
\hline
11$\times$9$\times$7         & $-$32.59(1)          & $-$22.329(4)  &  10.26(1)\\
22$\times$18$\times$14       & $-$32.01(2)          & $-$21.546(7)  & 10.46(2) \\
\end{tabularx}
\end{ruledtabular}
\label{table_H2_triplet}
\end{table}

\subsection{ H$_2$+H $\longrightarrow$ H+H$_2$ transition state} \label{sec:planewave_H3}

The problem of calculating the transition state of H$_3$
is well benchmarked using a variety of methods
\cite{H3_siegbahn,anderson_1994,anderson_2005,Peterson_H3,porezag_H3}.  
The activation energy for the reaction H$_2$+H $\longrightarrow$ H+H$_2$
is defined as the difference between the energy of the H$_3$ saddle point 
and that of the well separated H atom and H$_2$ molecule.

Density functional methods are not very accurate in calculating the
activation energy. For example, DFT with an LDA functional gives
H$_3$ as a bound molecule with a binding energy of $0.087$~eV at the
symmetric configuration with $R_1=R_2=1.795$\,Bohr.  DFT/GGA, on the
other hand, gives a barrier height of $0.152$~eV at the symmetric
configuration $R=1.767$\,Bohr \cite{porezag_H3}.
The experimental barrier height is 9.7 kcal/mol$ =0.420632$~eV \cite{H3_exp}.

\begin{table}[tb]
\caption{Symmetric collinear H$_3$ transition state energies using planewaves 
  with pseudopotentials. Results are shown from density
  functional GGA [with (PSP) and without (AE) pseudopotentials, respectively], DMC, and the present
  AF QMC methods. (The ``all-electron'' GGA(AE) results are from well converged large Gaussian basis set
  calculations.) The calculated results are for 
  the linear H$_3$ molecule with $R_1=R_2=R$, for three values of $R$. All energies are
  in eV. Statistical errors are on the last digit and are shown in
  parenthesis.}
\begin{ruledtabular}
\begin{tabular}{lcccc}
R           &  GGA(AE)  &  GGA(PSP)      &  DMC (exact)         & AF QMC      \\ 
\hline
$1.600$     & 0.297    &  0.30         & 0.543\,09(8)    & 0.54(3)  \\
$1.757$     & 0.156    &  0.16        & 0.416\,64(4)    & 0.43(3)  \\
$1.900$     & 0.222    &  0.22         & 0.494\,39(8)    & 0.48(4)  \\
\end{tabular}
\end{ruledtabular}
\label{table_H3_pw}
\end{table}

Using the AF QMC method, we studied the collinear H$_3$ system for three
configurations with R$_1$=R$_2$=1.600, 1.757, and 1.900\,Bohr. 
Table~\ref{table_H3_pw} shows the calculated barrier heights and compares
these to results from DFT/GGA all-electron and
pseudopotential calculations, and to results from recent DMC calculations \cite{anderson_2005}.
The AE and pseudopotential DFT/GGA results are in excellent
agreement with each other, a further indication of the good quality of
the H pseudopotential.  
The DMC calculations \cite{anderson_2005} are
exact in this case, through the use of a cancellation scheme \cite{FermionSign2,Anderson_cancel},
which is very effective at eliminating the sign 
problem for small systems.
The AF QMC values are in good agreement with 
the exact calculated results.

Planewave based AF QMC calculations of the H$_3$
transition state are very expensive, since the energy variations in 
the Born-Oppenheimer curve are quite small as seen in Table~\ref{table_H3_pw}.
To achieve the necessary accuracy, large supercells are needed, which results
in large planewave basis sets. The large basis sets lead to
many thousands of AF's in Eq.~(\ref{eq:HStrans1}).
Moreover, a large number of AF's in general lead to a more severe 
phase problem and thus potentially a more pronounced role for the phaseless approximation.
The larger AF QMC statistical errors, compared to the highly optimized 
DMC results as well as to our Gaussian basis results in 
Sec.~\ref{sec:Gaussian_H3}, reflect
the inefficiency of planewave basis sets for isolated
molecules. 
These calculations are valuable, despite their computational cost, as they 
demonstrate the robustness of the method.

\subsection{Hydrogen-bonded molecules}\label{sec:planewave_hbond}

Complementing the study above of the H$_3$ system, 
where energy differences are small,
we also examined three other
hydrogen-bonded molecules: H$_2$O, OH, and H$_2$O$_2$, where
the energy scales are large. 
Table~\ref{table_mol} compares the binding energies calculated
using DFT/GGA (both pseudopotential and all-electron), DMC \cite{grossman}, and the present
AF QMC method. 
(Results for the O$_2$ and O$_3$ molecules are included, because
they are pertinent to the discussion of pseudopotentials errors below.)
The experimental values \cite{feller_peterson}, 
with the zero point energy removed, are also shown.
All of the  calculations are performed
at the experimental geometries of the molecules.
The density functional all-electron binding energies in Table~\ref{table_mol} were obtained
using the highly converged triple-zeta ANO basis sets of Widmark, Malmqvist, and Roos
\cite{roos}. They are in good agreement with published
all-electron results. 
For example, the all-electron
binding energy of H$_2$O is 10.147~eV and that of OH is 4.77~eV in
Ref.~\cite{gga}. In Ref.~\cite{patton}, the binding energy of H$_2$O
is 10.265~eV, and that of O$_2$ is 6.298~eV \cite{patton}.

In all of the molecules except H$_2$O, the
DFT pseudopotential result seems to be in better agreement with the experimental value
than the all-electron result.  This is fortuitous and by no means
suggest that the pseudopotential results are better than the
all-electron values, since the pseudopotentials results should
reproduce the all-electron value obtained with the same theory. Any
differences are in fact due to transferability errors of the
pseudopotentials. 
At the density functional level, the molecular systems H$_2$O, OH, and
H$_2$O$_2$ all need a non-linear core correction (NLCC).  The NLCC was
introduced into DFT pseudopotential calculations by Louie \etal
\cite{nl_core}. It arises from the DFT-generated pseudopotential for
oxygen, at the pseudopotential construction level in the descreening
step, where the valence Hartree and nonlinear exchange-correlation
terms are subtracted to obtain the ionic pseudopotential. The Hartree
term is linear in the valence charge and can be subtracted exactly.
This is not the case with the nonlinear exchange-correlation
potential, and will lead to errors especially when there is an overlap
between the core and the valence charge densities.  According to the
NLCC correction scheme, this error can be largely rectified by
retaining an approximate pseudo-core charge density, and carrying it
properly in the target (molecular or solid) calculations.  This
generally improves the transferability of the pseudopotentials
\cite{nl_core,nl_core_porezag}.  The problem of NLCC is absent in
effective core-potentials (ECP) generated using the Hartree-Fock
method.

All of the molecules in Table~\ref{table_mol} suffer from the NLCC error which
originates predominantly from the spin-polarized oxygen atom, where
the NLCC can be as large as $0.3$~eV/atom within a GGA-PBE calculation
\cite{nl_core_porezag}. For this reason, we have also included
results for the O$_2$ and O$_3$ molecules. (The AF QMC value for O$_2$ is taken
Ref.~\cite{alsaidi_tio_mno}.) 
As seen in the table, the
binding energies of H$_2$O, OH, H$_2$O$_2$, O$_2$, and O$_3$ are smaller
than the corresponding all-electron values by $\approx 0.37$, $0.19$,
$0.60$, $0.50$, and $0.87$~eV, respectively. These values are
approximately proportional to the number of oxygen atoms in the
corresponding molecule with a proportionality constant $\approx
0.3$~eV, which agrees with the value reported in
Ref.~\cite{nl_core_porezag}.

The pseudopotential is of course used differently in many-body 
AF QMC calculations.
Despite the need for NLCC at the DFT level, the oxygen pseudopotential 
seems to be of good quality when used in AF QMC. 
In all the
cases, the  AF QMC results are in good agreement with DMC and with the experimental values. 
The largest
discrepancy with experiment is $\approx 0.4(2)$~eV with O$_3$, and it
is in opposite direction to the NLCC as done at the density functional
level.  

The need for the non-linear core correction does not indicate a failure of the
frozen-core approximation, but rather is a consequence of the non-linear dependence
of the spin-dependent exchange-correlation potential on the {\em total} spin-density
(valence+core) in density functional theory.
The QMC calculations depend only on the bare ionic pseudopotential and do not have this
explicit dependence on the (frozen) core-electron spin densities. 
It is thus reasonable to expect the QMC results to be not as sensitive to this issue.

\begin{table}[t]
  \caption{ 
    Calculated binding energies of H$_2$O, OH, H$_2$O$_2$, O$_2$, and
    O$_3$. Results are shown from density functional GGA [with (PSP) and without (AE) pseudopotentials, 
    respectively], DMC, and the present AF QMC methods. Experimental results are also shown.
    DFT/GGA(PSP) and the present AF QMC results were calculated using
    planewave basis sets with pseudopotentials. DFT/GGA(AE) is
    calculated using highly converged Gaussian basis sets. The DMC \cite{grossman}
    results were also obtained using pseudopotentials.
    The zero point energy is removed from the
    experimental data \cite{feller_peterson}. All energies are in
    eV. Statistical errors are on the last digit and are shown in
    parenthesis.}
\begin{ruledtabular}
\begin{tabularx}{3.7 in}{l ccccc}
           &GGA(AE)& GGA(PSP)&  DMC & AF QMC    & Expt.    \\
\hline
H$_2$O     & 10.19 &  9.82   & 10.10(8) & 9.9(1) & 10.09 \\
OH         & 4.79 &  4.60  & 4.6(1) & 4.7(1) & 4.63  \\
H$_2$O$_2$ & 12.26 & 11.66 & 11.4(1) &   11.9(3) & 11.65   \\
O$_2$      & 6.22 &  5.72  &        &  5.2(1) & 5.21  \\
O$_3$      & 7.99 &  7.12  &        & 6.2(2) & 5.82 \\
\end{tabularx}
\end{ruledtabular}
\label{table_mol}
\end{table}

\begin{table}[tb]
\caption{Symmetric collinear H$_3$ transition state total energies 
  using aug-cc-pVDZ and aug-cc-pVTZ Gaussian basis sets
  \cite{dunning1}. We examined 5 configurations with $R_1=R_2=R$, and
  we report the unrestricted Hartree-Fock (UHF), full configuration
  interaction (FCI), and AF QMC total energies.  Bondlengths are in
  Bohrs and energies are in Hartrees. Statistical errors are on the
  last digit and are shown in parenthesis. }
\begin{ruledtabular}
\begin{tabular}{l ddd}
\multicolumn{1}{c}{R } &\multicolumn{1}{c}{UHF} & \multicolumn{1}{c}{FCI}  & \multicolumn{1}{c}{AF QMC}     \\ 
\hline
\multicolumn{2}{l}{aug-cc-pVDZ}\\
 1.600 &  -1.595\,026 &  -1.642\,820  & -1.642\,56(5) \\
 1.700 &  -1.600\,252 &  -1.648\,186  & -1.647\,75(5)\\
 1.757 &  -1.601\,336 &  -1.649\,328  & -1.648\,82(5) \\
 1.800 &  -1.601\,406 &  -1.649\,433  & -1.648\,98(6)  \\
 1.900 &  -1.599\,536 &  -1.647\,606  & -1.646\,97(6)\\
\\
\multicolumn{2}{l}{aug-cc-pVTZ}\\
1.600 & -1.599\,843 & -1.652\,219  & -1.651\,78(7)    \\
1.700 & -1.604\,162 & -1.656\,405  & -1.655\,86(7)    \\
1.757 & -1.604\,835 & -1.657\,013  & -1.656\,52(7)   \\
1.800 & -1.604\,638 & -1.656\,770  & -1.656\,24(8)   \\
1.900 & -1.602\,269 & -1.654\,285  & -1.653\,68(9)  \\
\end{tabular}
\end{ruledtabular}
\label{table_H3bond}
\end{table}

\section{Results using Gaussian basis sets}\label{sec:gaussian_results}

In this section, we present our studies using
Gaussian basis sets. For comparison, some of 
 the systems are repeated from the
planewave and pseudopotential studies in the previous section. 
 Gaussian basis sets are in general more efficient for isolated molecules.
For example, the calculations below on the Van der Waals minimum in H$_3$
would be very difficult with the planewave formalism, 
because of the large supercells necessary, and because of the high
statistical accuracy required to distinguish the small energy scales.
Also, all-electron
calculations are feasible with a Gaussian basis, at least for lighter
elements, so 
systematic errors due to the use of pseudopotentials can be avoided without
incurring much additional cost.

Direct comparison with experimental results requires
large, well-converged basis sets in the AF QMC 
calculations \cite{gafqmc_postd,gafqmc}. As mentioned, the
convergence of Gaussian basis sets is not as straightforward to control
as that of planewaves. 
For benchmarking the accuracy of the AF QMC method, however, we can also compare 
with other established correlated methods such as
full configuration interaction (FCI) and CCSD(T),
since all the methods operate on the same 
Hilbert space.
FCI energies are the exact results for the 
Hilbert space thus defined. The FCI method has an exponential scaling with
the number of particles and basis size, so it is only used with small
systems. 
In this section, we
study H$_2$ and H$_3$, 
which are challenging examples for mean-field methods,
and compare the AF QMC results with exact results.

The matrix elements which enter in the definition of the Hamiltonian
of the system of Eq.~(1) are calculated using NWCHEM \cite{nwchem,
gafqmc}.  The trial wave functions, which are used to control the
phase problem, are mostly computed using unrestricted Hartree-Fock
(UHF) methods, although we have also tested ones from density
functional methods.  In previous studies, we have rarely seen any
difference in the AF QMC results between these two types of trial wave
functions.  This is the case for most of the systems in the present
work, and only one set of results are reported.  
In H$_3$ near the Van
der Waals minimum, where extremely small energy scales need to be resolved, 
we find small differences 
($\sim$ 0.1 milli-Hartree), and we report 
results from the separate trial wave functions. 
The FCI calculations
were performed using MOLPRO \cite{molpro,fci_molpro}.

\subsection{Bondlength of H$_2$}

We first study H$_2$ again, with a cc-pVTZ basis set which
has 28 basis functions for the molecule. This is to be
compared with the planewave calculations which has about
$5,000$ to $70,000$ planewaves for the different supercells used. 
These H-bonded systems are especially favorable for localized basis sets.
The AF QMC equilibrium bondlength
$R=1.4025(6)$\,Bohr
compares well to the corresponding FCI bondlength of $R=1.40265$\,Bohr, 
with both methods using the cc-pVTZ basis.
This is a substantially better estimate of the exact infinite 
basis result of $R_e=1.40083$\,Bohr \cite{roothaan} than was obtained from the 
planewave AF QMC results in Fig.~\ref{fig_H2bond}.
The remaining finite-basis error is much smaller than the 
statistical errors in the planewave calculations.
(The small residual finite-basis 
error is mostly removed at the cc-pVQZ basis set level,
with an equilibrium 
bondlength of $R=1.40111$\,Bohr
from FCI.)

\begin{table}[tb]
\caption{ H$_3$ total energies in
  the van der Walls limit.  $R_1$ is fixed at $1.4$\,Bohr, and $R_2$ is
  varied between 4 and 10 Bohr.  
The aug-cc-pVTZ basis set is used.
Energies are in Hartrees. Statistical errors
are on the last digit and are shown in parenthesis.}
\begin{ruledtabular}
\begin{tabular}{ldd}
\multicolumn{1}{c}{$R_2$} &   \multicolumn{1}{c}{FCI}   & \multicolumn{1}{c}{AF QMC/UHF} \\
\hline
4  &  -1.671\,577  & -1.671\,60(9)   \\
5  &  -1.672\,455  & -1.672\,50(8)   \\
6  &  -1.672\,535  & -1.672\,63(6)   \\
7  &  -1.672\,508  & -1.672\,65(5)   \\
10 &  -1.672\,462  & -1.672\,57(6)   \\
\end{tabular}
\end{ruledtabular}
\label{table_H3bond_vwalls}
\end{table}

\begin{figure}[t] 
\includegraphics[width=8.5cm]{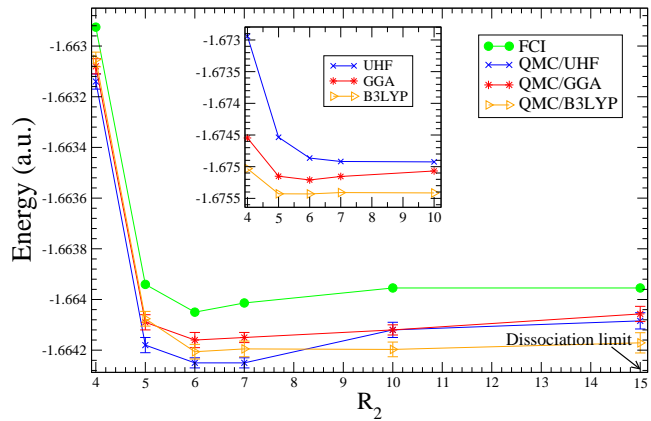} 
\caption{ 
Potential energy curve of H$_3$ in the van der Waals limit
  using aug-cc-pVDZ basis set.  $R_1$ is set to 1.4 Bohr, and $R_2$ is
  varied between 4 and 10 Bohrs. 
  (The dissociation limit is shown at  $R_2=15$\,Bohr in the figure). 
 FCI results are compared with AF QMC results with three different
  trial wave functions, from UHF and DFT with GGA and B3LYP functionals,
respectively. 
  The inset shows the corresponding potential
  energy curves obtained from
  UHF, GGA, and B3LYP.
  (For clarity,   the UHF
  and GGA energies are shifted by $-0.047$ and $-0.154$~Hartrees in the
  inset, respectively.)}
\label{fig_H3bond_vw}
\end{figure}

\subsection{ H$_2$+H $\longrightarrow$ H+H$_2$ transition state}\label{sec:Gaussian_H3}

Table~\ref{table_H3bond} presents calculated total energies of H$_3$
with aug-cc-pVDZ and aug-cc-pVTZ basis sets \cite{dunning1}. Results 
obtained with UHF, FCI, and the present AF QMC methods are shown for 
five different geometries in the collinear H$_3$ system. 
The present TZ-basis FCI results were cross-checked with those in
Ref.~\cite{Peterson_H3}, which contains a detailed study of the 
Born-Oppenheimer potential energy curves for the H+H$_2$ system.

The AF QMC total energies are in excellent agreement, to
within less than 1 mE$_H$, with the FCI energies.  The AF QMC
barrier heights with the aug-cc-pVDZ and aug-cc-pVTZ basis sets at
$R=1.757$\,Bohr are 0.444(2) and 0.434(3)~eV, respectively.  The
corresponding FCI results are 0.4309 and 0.4202~eV,
respectively. 
Thus the AF QMC results show a
systematic error of $\sim0.015$~eV in the barrier height.
It is possible to resolve these small discrepancies, 
because the basis sets are much more compact,
with $25$ to $75$ Gaussian basis functions as opposed to approximately 
$10,000$ planewaves in the calculations in Sec.~\ref{sec:planewave_H3}.
As a result, the statistical errors are 
smaller than in the planewave calculations by a factor of 10, 
with only a small fraction of the computational
time. Even with these relatively small basis sets,
we see that the finite-basis errors are
quite small here. In fact,
the FCI barrier height with the aug-cc-pVTZ basis is 
in agreement with the experimental value 0.420632~eV \cite{H3_exp}.

\subsection{Van der Waals minimum in collinear H$_3$ }\label{sec:Gaussian_VanderH3}

The van der Waals minimum of H$_3$ is studied by fixing $R_1=1.4$~Bohr 
(the  H$_2$ equilibrium bond length),
while the distance $R_2$ between the third H atom and
the closer of the two atoms in H$_2$ was varied between $4$ and $10$
Bohrs.
The potential energy curve of this system exhibits a very  
shallow minimum
of approximately $85\,\mu$E$_H$ \cite{Peterson_H3}
at $R_2\sim 6$\,Bohr. 
Two
different basis sets, aug-cc-pVDZ and aug-cc-pVTZ, were used.
Table~\ref{table_H3bond_vwalls} shows the aug-cc-pVTZ results,
and Fig.~\ref{fig_H3bond_vw} plots the
aug-cc-pVDZ results.

As seen in Table~\ref{table_H3bond_vwalls}, the AF QMC all-electron total
energies are in excellent agreement with FCI, with a maximum discrepancy of
about $0.14(5)$~mE$_H$.
The AF QMC energies, which are
calculated with the mixed-estimator, are not variational, as is evident in the results
from both basis sets compared to FCI. The AF QMC results in Table~\ref{table_H3bond_vwalls} 
are obtained with an UHF trial wave function. 
In most of our molecular calculations with Gaussian
basis sets, the UHF solution, which is the variationally 
optimal single Slater determinant, has  been 
chosen as the trial wave function \cite{gafqmc,gafqmc_postd}.
In the present case, the UHF method actually fails to
give a Van der Waals minimum, as can be seen from the inset of
Fig.~\ref{fig_H3bond_vw}. It is reassuring that AF QMC correctly reproduces the minimum 
with UHF as a trial wave function.

The effects of using two other single Slater determinant trial wave functions were also
tested. These were obtained from
DFT GGA and B3LYP calculations with the aug-cc-pVDZ basis set.
The corresponding results are also shown in
Fig.~\ref{fig_H3bond_vw}.
In the DFT calculations (shown in the inset of Fig.~\ref{fig_H3bond_vw}), 
both GGA and B3LYP predict the existence of a minimum, although  
B3LYP gives an unphysical small barrier at about $R_2\approx 7$\,Bohr. 
The AF QMC results 
obtained with UHF, GGA, and B3LYP Slater determinants as trial wave functions 
differ somewhat, but 
are reasonably close to each another.
With the GGA trial wave function, AF QMC ``repairs'' the well depth (possibly with 
a slight over-correction). With the B3LYP trial wave function, AF QMC appears to 
under-estimate the well-depth, giving a well-shape that is difficult to 
characterize because of the statistical errors
and the extremely
small energy scale of these features.

\section{Summary}
\label{sec:summary}

We have presented a benchmark study of the phaseless AF QMC method in 
various H-bonded molecules.
The auxiliary-field QMC method is a many-body approach formulated in a
Hilbert space defined by a single-particle basis. The choice of a
basis set is often of key importance, as it can affect 
the efficiency of the calculation. In the case of AF QMC, 
the basis set choice can also affect 
the systematic error, because of the 
different HS transformation that can result.
In this study, we 
employed planewave basis sets with pseudopotentials and all-electron
Gaussian basis sets, 
to compare the performance of the AF QMC method.
The planewave HS decomposition was tailored
to the planewave representation, resulting in ${\mathcal O}(8~M)$ auxiliary fields, where $M$ is the number
of planewaves. For the Gaussian basis sets, the generic HS decomposition described in 
Section~\ref{sec:AFQMC-method} was used, resulting in ${\mathcal O}(M^2)$ auxiliary
fields. 
Typical $M$ values in this study were tens of thousands in the planewave
calculations and a hundred in the Gaussian calculations.

The planewave calculations were carried out for H$_2$, ${\rm H}_2+{\rm H}$ near
the transition state, H$_2$O, OH, and H$_2$O$_2$. Non-linear core corrections to the oxygen
pseudopotential were discussed using additional calculations for the O$_2$
and O$_3$ molecules. 
DFT GGA pseudopotentials were employed. 
The trial wave functions were single Slater
determinants obtained from DFT GGA with identical planewave and pseudopotential 
parameters as in the AF QMC calculations. 
Hard pseudopotentials and large planewave cutoffs were used to ensure basis-size convergence and the
transferability of the pseudopotentials. 
Large supercells were employed to
remove finite-size errors. 
To mimic typical systems in the solid state, no optimization was done to 
take advantage of the simplicity 
of these particular systems. 
The binding energies computed from AF QMC have statistical errors of 
0.1-0.3\,eV as a result. Within this accuracy, the AF QMC results are 
in excellent agreement with 
experimental values. 

Gaussian basis AF QMC calculations were carried out on  H$_2$,
the transition state of ${\rm H}_2+{\rm H}$, as well as the 
van der Waals minimum in linear ${\rm H}_2+{\rm H}$. These 
calculations are within the framework of standard quantum chemistry many-body
using the full Hamiltonian without pseudopotentials. 
UHF single Slater determinants were used as the trial wave 
function. 
For various geometries, the absolute total energies from AF QMC agree with FCI
to well within 1 mE$_H$. The calculated equilibrium bondlengths and 
potential energy curves are also in excellent agreement with FCI. 
In ${\rm H}_2+{\rm H}$, AF QMC correctly recovers
the van der Waals well with a UHF trial wave function which in itself predicts
no binding.

Comparing planewave and Gaussian basis set AF QMC results, we can
 conclude the following.  In the Gaussian basis calculations, as
 evident from FCI comparisons, errors due to controlling the phase
 problem in the phaseless approximation are well within 1 mE$_H$ in
 the absolute energies.  Achieving the infinite basis limit is more
 straightforward using planewave based AF QMC, but statistical errors
 are larger for the isolated molecules studied due to the need for
 large supercells. Within statistical errors, however, the AF QMC
 results using both types of basis sets were in agreement. This
 indicates that errors due to the use of pseudopotentials with
 planewave basis sets were smaller than the statistical
 errors. Finally, within statistical errors, the performance of the
 phaseless AF QMC method, did not appear to be sensitive to the type
 of HS decomposition used, despite drastic differences in basis size
 and the number of auxiliary fields.  
 was tailored

\section{Acknowledgments:}

We would like to thank E.~J.~Walter for many useful discussions.
This work is supported by ONR (N000140110365 and N000140510055),
NSF (DMR-0535529), and ARO (grant no.~48752PH).  
Computations were carried out in part at the
Center for Piezoelectrics by Design, the SciClone Cluster at the
College of William and Mary, and NCSA at UIUC supercomputers.


\begin{thebibliography}{10}
\bibitem{QMC_rmp} W.~M.~C.~Foulkes, L.~Mitas, R.~J.~Needs, and
G.~Rajagopal, Rev. Mod. Phys. {\bf 71}, 33 (2001).

\bibitem{zhang_krakauer} S. Zhang and H. Krakauer, Phys. Rev. Lett. {\bf 90},
  136401 (2003).

\bibitem{gafqmc} W. A. Al-Saidi, S. Zhang, and H. Krakauer,
  J. Chem. Phys. {\bf 124}, 224101(2006).


\bibitem{payne_rmp}{ M. C. Payne, M. P. Teter, D. C. Allan, T. A. Arias
and J. D. Joannopoulos,  Rev. Mod. Phys. {\bf 64}, 1045 (1992).}

\bibitem{basis_sets_web} A compilation of basis sets is present at the
  Extensible Computational Chemistry Environment Basis Set Database
  (http://www.emsl.pnl.gov/forms/basisform.html).

\bibitem{dunning1} { T. H. Dunning, Jr., J. Chem. Phys. {\bf 90}, 1007
(1989).}

\bibitem{dunning2} { D.~E. Woon and T. H. Dunning, Jr,
J. Chem. Phys. {\bf 98}, 1358 (1993).}

\bibitem{FermionSign} {D. M. Ceperley and B. J. Alder, J. Chem. Phys. {\bf 81}, 5833 (1984);
 J. B. Anderson in {\it Quantum Monte Carlo: Atoms, Molecules,\ Clusters, Liquids and Solids},
 Reviews in Computational Chemistry, Vol. 13, ed. by Kenny B. Lipkowitz and
 Donald B. Boyd (1999).}

\bibitem{FermionSign2} {Shiwei Zhang and M. H. Kalos, Phys. Rev. Lett. {\bf 71}, 2159 (1993)}

\bibitem{Zhang} Shiwei Zhang, J.~Carlson, and J.~E.~Gubernatis, Phys.\
Rev.\ B {\bf 55}, 7464 (1997).

\bibitem{anderson_fixednode} J.~B.~Anderson, J. Chem. Phys. {\bf 63},
  1499 (1975).

\bibitem{dmc} J. W. Moskowitz, K. E. Schmidt, M. A. Lee, and
Malvin H. Kalos, J.\ Chem.\ Phys. {\bf 77}, 349 (1982);
Peter J. Reynolds, David M. Ceperley, B. J. Alder, and W. A. Lester,
J.\ Chem.\ Phys. {\bf 77}, 5593 (1982). 



\bibitem{alsaidi_tio_mno} W. A. Al-Saidi, H. Krakauer, and S. Zhang, Phys. Rev. B
{\bf 73}, 075103 (2006).
  

\bibitem{gafqmc_postd} W. A. Al-Saidi, H. Krakauer, and S. Zhang,
J. Chem. Phys. {\bf 125}, 154110 (2006).


\bibitem{gafqmc_bond} W. A. Al-Saidi, H. Krakauer, and S. Zhang, in
  preparation. 

\bibitem{HS} R.~L.~Stratonovich, Sov.\ Phys.\ Dokl. \textbf{2},
416 (1958); J.~Hubbard, Phys.~Rev.~Lett.
{\bf 3}, 77 (1959).

\bibitem{nwchem}{
 T.~P.~Straatsma, E.~Apr\'a, T.~L.~Windus, E.~J.~Bylaska, W.~de Jong,
            S.~Hirata, M.~Valiev, M.~T.~Hackler, L.~Pollack, R.~J.~Harrison,
            M.~Dupuis, D.~M.~A.~Smith, J.~Nieplocha, V.~Tipparaju,
            M.~Krishnan, A.~A.~Auer, E.~Brown, G.~Cisneros, G.~I.~Fann,
            H.~Fruchtl, J.~Garza, K.~Hirao, R.~Kendall, J.~A.~Nichols,
            K.~Tsemekhman, K.~Wolinski, J.~Anchell, D.~Bernholdt, P.~Borowski,
            T.~Clark, D.~Clerc, H.~Dachsel, M.~Deegan, K.~Dyall, D.~Elwood,
            E.~Glendening, M.~Gutowski, A.~Hess, J.~Jaffe, B.~Johnson, J.~Ju,
            R.~Kobayashi, R.~Kutteh, Z.~Lin, R.~Littlefield, X.~Long, B.~Meng,
            T.~Nakajima, S.~Niu, M.~Rosing, G.~Sandrone, M.~Stave, H.~Taylor,
            G.~Thomas, J.~van Lenthe, A.~Wong, and Z.~Zhang,
            ``NWChem, A Computational Chemistry Package for Parallel Computers, 
            Version 4.6'' (2004),
                      Pacific Northwest National Laboratory,
                      Richland, Washington 99352-0999, USA.}


\bibitem{roothaan} W. Kolos and C. C. J. Roothaan, Rev. Mod. Phys. {\bf
  32}, 219 (1960). 


\bibitem{rappe} Andrew~M.~Rappe, Karin M. Rabe, Efthimios Kaxiras, and
  J.~D.~Joannopoulos, Phys.\ Rev.\ B {\bf 41}, R1227 (1990).

\bibitem{CPC05} Shiwei Zhang, Henry Krakauer, Wissam Al-Saidi, and
Malliga Suewattana, Comp. Phys. Comm. {\bf 169}, 394 (2005).

\bibitem{cherry} Malliga Suewattana, W. Purwanto, S. Zhang,
  H. Krakauer, and E.~J.~Walter (unpublished). 

\bibitem{nl_core} Steven G. Louie, Sverre Froyen, and Marvin L. Cohen 
  Phys. Rev. B {\bf 26}, 1738 (1982).

\bibitem{nl_core_porezag} Dirk Porezag, Mark R. Pederson, and Amy Y. Liu ,
  Phys. Rev. B {\bf 60}, 14132 (1999).

\bibitem{gga} J.~P.~Perdew, K.~Burke, and M.~Ernzerhof, Phys.  Rev. Lett
 {\bf 77}, 3865 (1996).

\bibitem{patton} David C. Patton, Dirk V. Porezag, and Mark
 R. Pederson, Phys. Rev. B {\bf 55}, 7454 (1996).


\bibitem{Silvestrelli93} P. L. Silvestrelli, S. Baroni, and R. Car, 
Phys.\ Rev.\ Lett.\ {\bf 71}, 1148 (1993). 

\bibitem{H3_siegbahn} P. Siegbahn and B. Liu, J.  Chem. Phys. {\bf 68},
  2457 (1978).

\bibitem{anderson_1994} Drake L. Diedrich and James B. Anderson, J.
  Chem. Phys. {\bf 100}, 8089 (1994).

\bibitem{anderson_2005} Kevin  E. Riley  and James B. Anderson, J.
  Chem. Phys. {\bf 118}, 3437 (2002).

\bibitem{Peterson_H3} Steven L. Mielke, Bruce C. Garrett, and
Kirk~A.~Peterson, J.~Chem.~Phys. {\bf 116}, 4142 (2001).

\bibitem{porezag_H3} Dirk Porezag and Mark R. Pederson,
  J. Chem. Phys. {\bf 102}, 9345 (1995).

\bibitem{H3_exp} W. R. Schulz and D. J. Le Roy, J. Chem. Phys. {\bf42}, 3869 (1965).

\bibitem{Anderson_cancel} J. B. Anderson, C. A. Traynor, and
  B.~M.~Boghosian, J. Chem. Phys. {\bf 95}, 7418 (1991).

\bibitem{grossman} Jeffery C. Grossman, J. Chem. Phys. {\bf 117}, 1434
(2002).

\bibitem{feller_peterson} David Feller, and Kirk A. Peterson,
  J. Chem. Phys. {\bf 110}, 8384 (2002).

\bibitem{roos} {P. O. Widmark, P. A. Malmqvist, and B. Roos,
Theo.~Chim.~Acta, 77, 291 (1990).}

\bibitem{molpro} MOLPRO version 2002.6 is a package of ab initio programs written by
H.-J.~Werner and P.~ J.~Knowles, with contributions from J.~Almlof,
R.~D.~Amos, M.~J.~O.~ Deegan, S.~T.~Elbert, C.~Hampel, W.~Meyer,
K.~A.~Peterson, R.~M.~Pitzer, ¨ A.~J.~Stone, P.~R.~Taylor, and
R.~Lindh, Universitat Bielefeld, Bielefeld, Germany, University of
Sussex, Falmer, Brighton, England, 1996.


\bibitem{fci_molpro} {P.~J.~Knowles and N.~C.~Handy,
  Chem.~Phys.~Letters 111, 315 (1984); P.~J.~Knowles and N.~C.~Handy,
  Comp.~Phys.~Commun.~54, 75 (1989).}


\end{thebibliography}
\end{document}